\documentclass[apjl]{emulateapj}
\usepackage{natbib}
\usepackage{hyperref}
\usepackage{graphicx}

\shorttitle{X-ray Measurement of Solar Max}
\shortauthors{Winter \& Balasubramaniam}

\DeclareFixedFont{\BI}{OT1}{ptm}{b}{it}{11}

\begin{document}
%
%


\title{Estimate of Solar Maximum using the 1-8 \AA~Geostationary Operational Environmental Satellites X-ray Measurements}  
\vspace{0.5cm}

\author{L.~M. Winter}
\affil{Atmospheric and Environmental Research, 131 Hartwell Avenue, Lexington, MA 02421, USA}
\email{lwinter@aer.com}
\and
\author{K. S. Balasubramaniam}
\affil{Space Weather Effects Section, Space Vehicles Directorate, Air Force
Research Laboratory, Kirtland AFB, NM 87117, USA}

\begin{abstract} 

We present an alternate method of determining the progression of the solar cycle through an analysis of the solar X-ray background.  Our results are based on the NOAA Geostationary Operational Environmental Satellites (GOES) X-ray data in the 1-8 \AA~ band from 1986 - present, covering solar cycles 22, 23, and 24.  The X-ray background level tracks the progression of the solar cycle through its maximum and minimum.  Using the X-ray data, we can therefore make estimates of the solar cycle progression and date of solar maximum. Based upon our analysis, we conclude that the Sun reached its hemisphere-averaged maximum in Solar Cycle 24 in late 2013. This is within six months of the NOAA prediction of a maximum in Spring 2013.  
\end{abstract}

\section{Introduction}
	Predictions of the length of the solar cycle and date of solar maximum are important for planning space missions and satellite orbits.  Besides direct electromagnetic, particle, and mass effects, the sun cyclically influences the terrestrial ionospheric structure and interplanetary structure.  A number of empirical or semi-empirical methods for estimating solar cycle progression exist.  The earliest used method relies upon the sunspot number, following the discovery by \citet{1852MNRAS..13...29W}
of the 11-year periodicity in sunspot activity.  The "geomagnetic precursor'' methods, relying upon measurements of changes in the Earth's magnetic field, determine correlations between sunspot number at solar maximum and the geomagnetic $aa$ index at the preceding minimum (e.g.,	\citealt{1979stp.....2..258O, 1982JGR....87.6153F,1993SoPh..148..383T}).  Additionally, the solar radio emission at 10.7 cm (F10.7) is a consistent measurement that has been recorded daily since 1947 and is also found to follow the solar activity cycle \citep{1982JGR....87.6153F}.

	Combinations of these techniques have been used to predict the intensity and date of solar maximum of the current solar cycle.  The Solar Cycle 24 Prediction Panel\footnote{The Consensus statement of the solar cycle 24 prediction panel is available at \url{http://www.swpc.noaa.gov/SolarCycle/SC24/}.} \citep{2007AAS...210.9206B}, led by NOAA, examined several techniques and predicted a maximum in May 2013 that would be weak compared to recent solar cycles.  Similarly, recent work presented in \citet{2014SoPh..289.2317P} predicts Solar Cycle 24 maximum F10.7 of no stronger than average and likely weaker than recent solar cycles.

	In this letter, we present a novel approach for determining the solar cycle peak and duration.  The solar X-ray background, like other tracers such as the sunspot number and solar radio emission, rises during active times and declines in quiet times.  Through an analysis of the X-ray data from the past few solar cycles, we predict the maximum X-ray background level, date of solar maximum, and length of Solar Cycle 24.  In Section 2, we describe our analysis.  In Section 3, we compare the X-ray background results to the monthly sunspot number.  Section 4 includes discussion of our results.

\section{Determination of the Solar Cycle Maximum through the X-ray Background}
To make our solar cycle predictions, we analyzed GOES X-ray observations obtained from NOAA's NGDC\footnote{The GOES SEM data are available at: \url{http://www.ngdc.noaa.gov/stp/satellite/goes/dataaccess.html}}.  We determined the 1-8\,\AA~(corresponding to $\sim 1.5 - 12.4$\,keV) background levels using 1-minute data from 1986 through May 15, 2014.  The data were obtained from GOES-6, -7, and -8 (solar cycle 22), GOES-8 and -10 (solar cycle 23), and GOES-14 and -15 (2009 -- present).   

The X-ray background was computed as the smoothed minimum flux in a 24-hr time period preceding each 1-minute GOES observation.  In detail, we use the technique of \citet{SWE:SWE20042},
which includes the following steps: (1) compute the hourly median with a sliding 1-hour window, (2) determine the instantaneous background as the minimum of these hour medians in the previous 24 hours, and (3) smooth the instantaneous background by the previous 2 hours.  The background was computed for both the 1-4\,\AA~and 1-8\,\AA~GOES observations.  The harder X-ray emission shows no discernible solar cycle trends, when compared with the soft X-ray emission, which is the focus of this paper.

In order to determine the solar maximum and length of solar cycle for cycles 22-24, we fit a simple Gaussian to the X-ray background of each solar cycle.  We chose a Gaussian for its simplicity in requiring only three free parameters and for its ability to reproduce the shape of the data over a solar cycle.  To fit the data, we converted the date and time of the observation into decimal years from the start of the solar cycle ({\BI SCY}).  We identified Solar Cycle 22 as beginning in August 1986 and ending by May 1996; Solar Cycle 23 as beginning in May 1996 and ending by December 2008; and Solar Cycle 24 as beginning in December 2008.  We then fit a Gaussian of the form: 
\begin{equation}
F({\rm SCY}) = F_0 \exp(({\rm SCY} - Solar\,Max)^2/(2 \sigma^2)), 
\end{equation}
to the X-ray background.  In the equation, {\BI F} is the logarithm of the X-ray background flux in W m$^{-2}$, {\BI F$_0$} is the logarithm of the X-ray background flux at solar maximum in W m$^{-2}$, {\BI SCY} is the solar cycle year in years, {\BI Solar Max} is the fitted solar maximum value in years from the start of the solar cycle, and $\BI{ \sigma}$ is the half-width of the solar cycle.  In the fitting process, we filtered out any data points with background levels below $10^{-9}$\,W\,m$^{-2}$.  Such measurements are below the GOES 1-8\,\AA~threshold of $3.7 \times 10^{-9}$\,W\,m$^{-2}$.  The Levenberg-Marquardt algorithm \citep{cite-levenberg,cite-marquardt}
was used to find the best-fit parameters {\BI F$_0$}, {\BI Solar Max}, and $\sigma$ with the SciPy optimization library in Python.

\begin{figure}
\begin{center}

\includegraphics[width=4in]{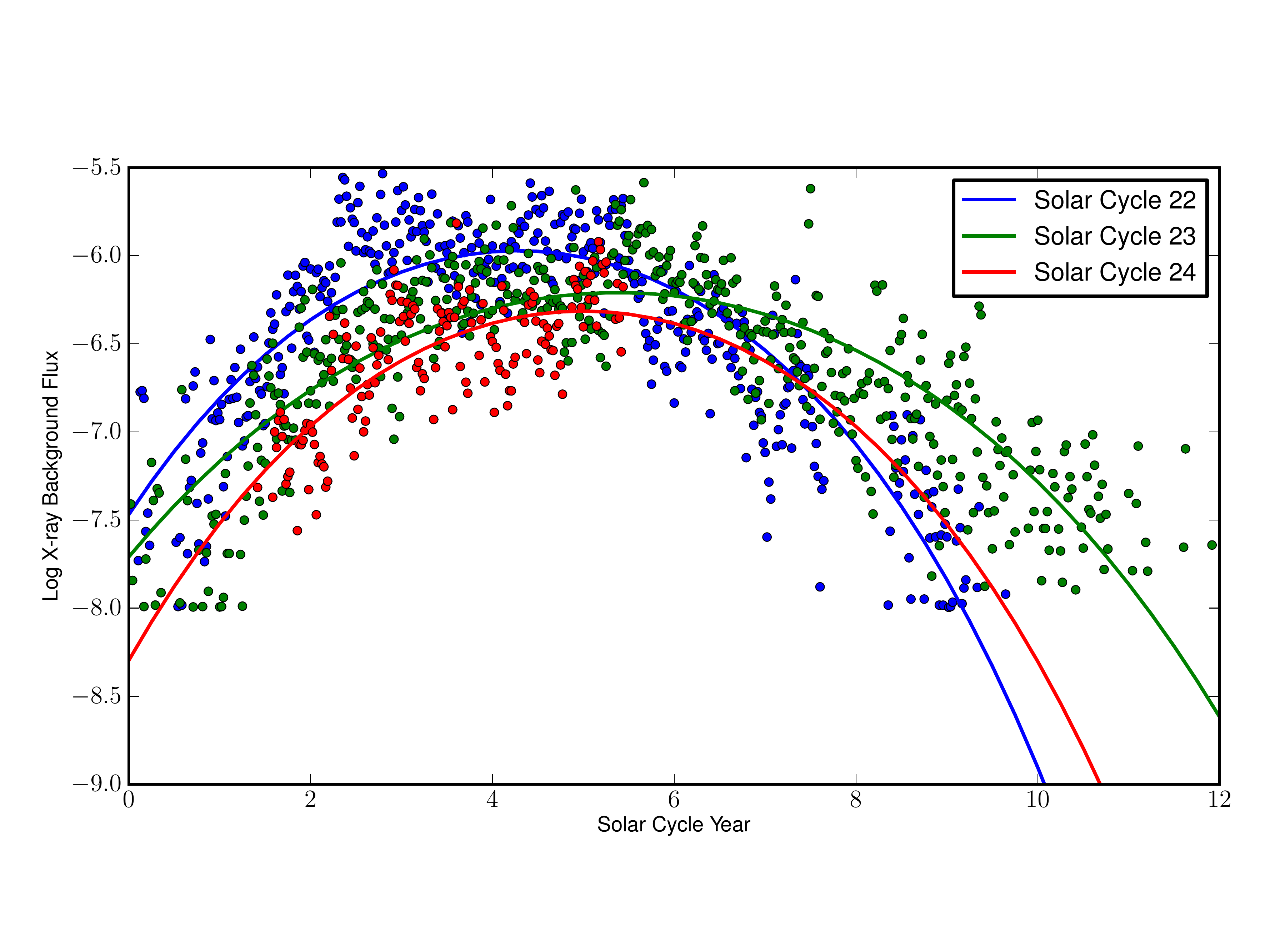}

\caption{Results of a Gaussian fit (lines) to the 1-week averaged 1-8 \AA~X-ray background (points; in units of log of W m$^{-2}$) for Solar Cycles 22, 23, and 24.  The X-ray background flux varies within the solar cycle, with higher values by a factor of 100 from solar minimum to solar maximum.}\label{fig-gaussian}
\end{center}
\end{figure}

We determined the effect of the choice of bin size on the solar cycle parameters by computing best-fit values and $\chi^2$ statistics for bin sizes of 1 month, 2 weeks, 1 week, 2.5 days, and 1 day.  The best-fit parameters, for each solar cycle examined, and for each binning level are listed in Table~\ref{table-binning}.  We find that the peak background flux is the most stable parameter, with very little variation in this parameter regardless of bin size.  For Solar Cycles 22 and 23 the solar maximum calculation is also stable, but the duration of the cycle varies by 2 months for cycle 22 and 6 months for cycle 23.  We tested goodness of fit with the chi-squared statistic, defined as \begin{math} \chi^2 = \Sigma ({\rm observed} - {\rm model})^2 / {\rm std}^2\end{math}, where std is the standard deviation of the measurements.  The ideal case is where the reduced $\chi^2$ statistic, $\chi^2$ divided by the degrees of freedom (the number of data points fitted minus the number of free parameters fit by the model), is closest to one.  For the smaller bin sizes, cases where the reduced $\chi^2$ value is much greater than one are labeled as {\it oversampled}.  For the largest bin sizes, the standard deviation is large, causing reduced $\chi^2 << 1$.  The optimized reduced $\chi^2$ values in Table~\ref{table-binning} correspond to the 1-week binning.  The best-fit parameters from the 1-week bin size are shown in Table~\ref{table-bestfit}.  The median 1-week background and best-fit Gaussians are shown for solar cycles 22, 23, and 24 in Figure~\ref{fig-gaussian}.
	   
Traditional measures of the solar cycle such as sunspot numbers show a double-peak due to the solar activity in the northern and southern hemispheres (e.g., \citealt{1977SoPh...52...53R}).
Similarly, the X-ray observations also show the double peak profile.  However, our choice of binning size affects whether the double peaked structure is blurred or distinct. For this reason we chose to fit only a single Gaussian to derive the solar maximum and duration, but determined the peaks from examination of the 1-week binned data.  In  Table~\ref{table-bestfit}, Peak 1 corresponds to the peak in the X-ray background occurring before the fitted solar maximum and Peak 2 is the peak following the solar maximum.

	    Since the current solar cycle 24 is incomplete, the resulting fewer measurements lead to  more variability in the fitted solar maximum and duration parameters depending on the chosen bin size.  In all cases (Table 2), however, we find that we have reached or passed the solar maximum.  The solar cycle 24 is likely to end around 2020, with a maximum uncertainty of 2 years.

\begin{figure}
\begin{center}

\includegraphics[height=2.5in]{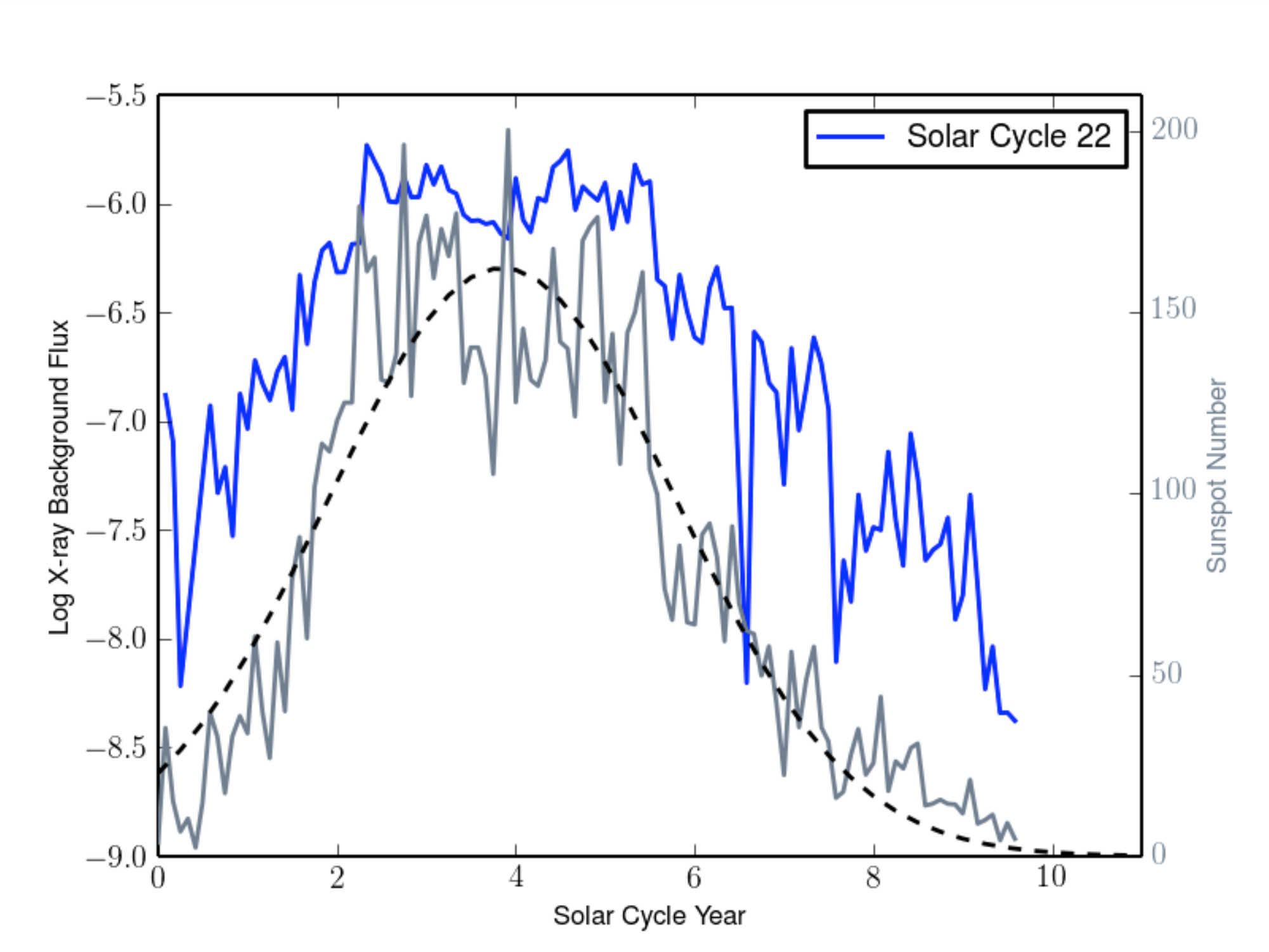}

\includegraphics[height=2.5in]{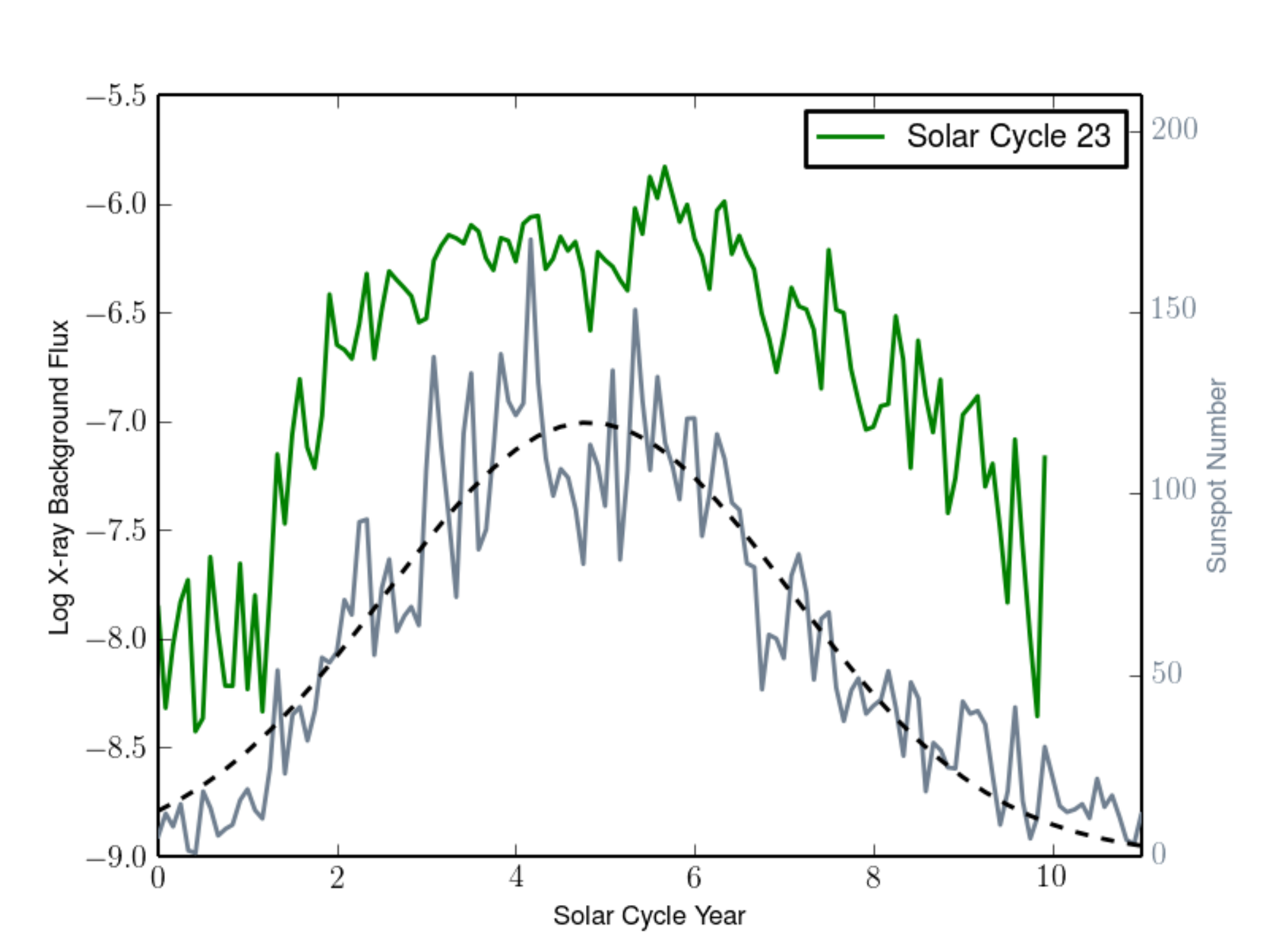}

\includegraphics[height=2.5in]{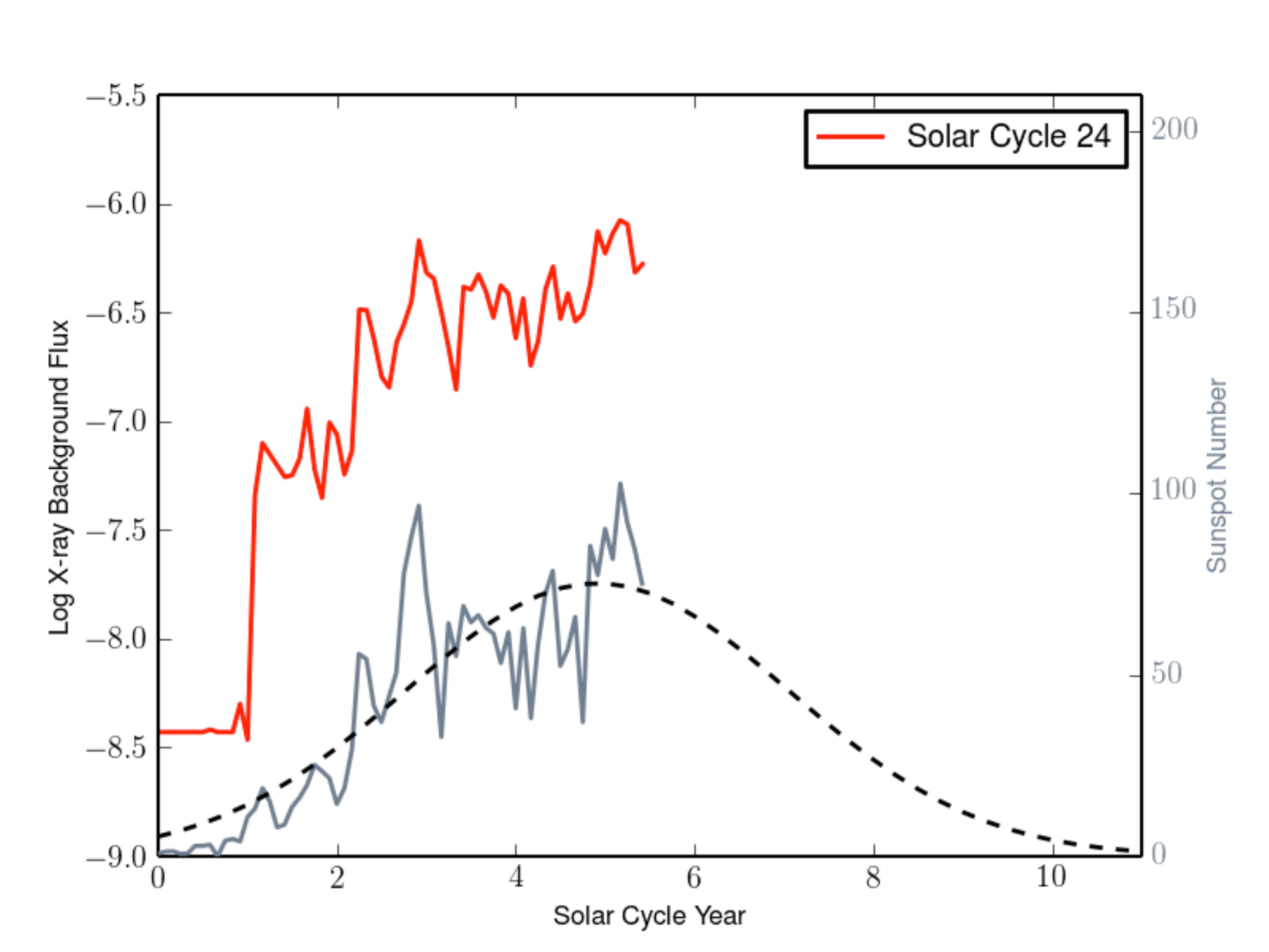}

\caption{Comparison of the 1-month averaged 1-8 \AA~X-ray background (colored lines; in units of log of W m$^{-2}$) to the monthly sunspot number (gray line) for Solar Cycles 22, 23, and 24. Dashed lines show the Gaussian fit to the sunspot number.  Note that color notations are the same as in Figure 1.}\label{fig-ssn}
\end{center}
\end{figure}
	
\section{Comparison to the Sunspot Number}\label{sunspot}
The earliest used method of determining the solar activity level was through observing changes in the sunspot number.  To determine how the X-ray background compares to sunspot number, we analyzed data of the monthly sunspot number from the Solar Influences Data Analysis Center in Belgium\footnote{\small The monthly sunspot number was obtained from NASA Marshall Space Flight Center's compilation available here: \url{http://solarscience.msfc.nasa.gov/greenwch/spot_num.txt}}.  In Figure~\ref{fig-ssn}, we show the sunspot number on the same scale as the 1-month averaged 1-8 \AA~X-ray background, for solar cycles 22-24.  We fit gaussians to the sunspot number data for each of the Solar Cycles 22-24, with the best-fit parameters listed in Table~\ref{table-sunspot}.


In addition to the best-fit solar maximum date and associated sunspot number, Table~\ref{table-sunspot} also includes the date and sunspot number of each of the peaks, obtained from analysis of the sunspot number data.  Peak 1 is the peak sunspot number before solar maximum and Peak 2 is the peak sunspot number following solar maximum.  Given the difference in the binning method for the sunspot number from the X-ray background, we expect some variability in comparing the dates of solar max.  We do find reasonable agreement, however, even when comparing with the 1-week binning of the 1-8 \AA~X-ray background (Table~\ref{table-bestfit}).  For cycles 22 and 23, the X-ray background solar maximum (regardless of binning) is within about 4 months from the sunspot number solar maximum.  The cycle 24 estimates agree for peak 2 (Feb 2014; which is the peak to date of the analysis), which was the maximum in sunspot number to date, although there is a large difference in the peak 1 dates of 8 months.

The peaks in the sunspot number and X-ray background are both higher preceding the solar max for
cycles 22 and 23.  The sunspot peak numbers in cycle 22, 192 for peak 1 and 173 for peak 2, were higher than in subsequent cycles.  The overall highest peaks in sunspot number from cycles 23 and 24 are 13\% and 48\% lower than the peak in cycle 22.


\section{Discussion}
The solar soft X-ray emission is an important indicator of the state of the corona.  While the mechanisms of coronal heating are poorly understood, the process is connected with solar magnetic activity (e.g., \citealt{1978ARA&A..16..393V}). Previous soft X-ray studies have shown that variations exist in the derived luminosity from minimum to maximum, by a ratio of 5 to 6 times \citep{2003ApJ...593..534J}. With uniform observations over the past nearly three solar cycles, the GOES soft X-ray measurements provide a powerful database for characterizing the coronal variability and a tool for not only monitoring of flare activity but also for space weather forecasting.

Based on our analysis of the GOES 1-8\,\AA~($\sim$ 1.5-12\,keV) observations from 1986--present, we have confirmed that the X-ray emission varies with solar cycle.  We determined a soft X-ray background as the minimum flux in a 24-hr time period preceding each 1-minute GOES observation.  From our analysis, we show that the variance in this X-ray background follows a cyclical pattern from solar minimum to maximum.  Additional variations between solar cycles (e.g., differences in the solar maximum flux and length of the cycle) are also found, with the peak background at solar maximum declining over the past two cycles. In particular, we find that the Solar Cycle 22 X-ray background peak of $1.1 \times 10^{-6}$ W m$^{-2}$ is 1.6 times the Solar Cycle 23 peak.  The predicted peak for Solar Cycle 24 is $5.21 \times 10^{-7}$ W m$^{-2}$, 25\% lower than the peak background level in cycle 23 and only half of the peak level in cycle 22.  This variance is consistent with the variability found in the sunspot cycle during the same time periods, as shown in \S~\ref{sunspot}. 

Further, we find that the soft X-ray emission during solar minimum has also declined over the past two cycles.  The average level over the year of solar minimum preceding each solar cycle declined by a factor of $\sim 4$ from  $8.3 \times 10^{-8}$ W m$^{-2}$ during Solar Cycle 22 to $2 \times 10^{-8}$ W m$^{-2}$ during Solar Cycle 23.  During Solar Cycle 24, the solar minimum average is unable to be determined reliably, since 72\% of the measurements in solar minimum were below the GOES threshold of $3.72 \times 10^{-8}$ W m$^{-2}$.  However, this evidence shows that the background was lower than the previous minimum, consistent with results from more sensitive soft X-ray instruments such as the SphinX X-ray spectrophotometer on the Russian CORONAS-PHOTON spacecraft (e.g., \citealt{2012ApJ...751..111S}).

These results carry merit as a historical study of the solar X-ray emission. Additionally, they exhibit the potential of our technique for space weather climatology.  One use is as an alternative method in determining the characteristics of the solar cycle.  Based upon our results, we predict the hemisphere-averaged maximum for Solar Cycle 24 as occurring in Nov 2013, with the peak so far having occurred in Feb 2014.  The X-ray based predictions we made for the previous two solar cycles, along with the current cycle, were in good agreement with the sunspot cycle.  Our analysis also allows us to estimate the end date of the current cycle.  Our predicted end dates for Solar Cycles 22 and 23 were April 1997 and Nov 2009, respectively.  In cycle 23, our predicted end date is $\sim$1 year later than NOAA SWPC's agreed upon date of Dec 2008.  Our predicted end date for Solar Cycle 24 is Sep 2020.

Additionally, since the occurrence of X-ray flares are linked to solar activity and we have shown that the soft X-ray background scales with this, the X-ray background may also prove an important tool in X-ray flare forecasting.  The intensity and number of flares, for instance, is also shown to scale with solar cycle.  In future work, we will explore the use of this technique as a diagnostic for {\it in-progress} flare forecasting. We also plan to compare these solar cycle measures to those from observations with the coronagraph at the John W. Evans Solar Facility of the National Solar Observatory at Sacramento Peak  \citep{2014SoPh..289..623A}.


\begin{thebibliography}{14}
\expandafter\ifx\csname natexlab\endcsname\relax\def\natexlab#1{#1}\fi

\bibitem[{{Altrock}(2014)}]{2014SoPh..289..623A}
{Altrock}, R.~C. 2014, \solphys, 289, 623, 623

\bibitem[{{Biesecker} \& {Prediction Panel}(2007)}]{2007AAS...210.9206B}
{Biesecker}, D.~A., \& {Prediction Panel}, S.~C.~. 2007, in Bulletin of the
  American Astronomical Society, Vol.~39, American Astronomical Society Meeting
  Abstracts \#210, 209

\bibitem[{{Feynman}(1982)}]{1982JGR....87.6153F}
{Feynman}, J. 1982, \jgr, 87, 6153, 6153

\bibitem[{Hock {et~al.}(2013)Hock, Woodraska, \& Woods}]{SWE:SWE20042}
Hock, R.~A., Woodraska, D., \& Woods, T.~N. 2013, Space Weather, 11, 262, 262

\bibitem[{{Judge} {et~al.}(2003){Judge}, {Solomon}, \&
  {Ayres}}]{2003ApJ...593..534J}
{Judge}, P.~G., {Solomon}, S.~C., \& {Ayres}, T.~R. 2003, \apj, 593, 534, 534

\bibitem[{{Levenberg}(1944)}]{cite-levenberg}
{Levenberg}, K. 1944, Quarterly of Applied Mathematics, 2, 164, 164

\bibitem[{{Marquardt}(1963)}]{cite-marquardt}
{Marquardt}, D. 1963, SIAM Journal on Applied Mathematics, 11, 431, 431

\bibitem[{{Ohl} \& {Ohl}(1979)}]{1979stp.....2..258O}
{Ohl}, A.~I., \& {Ohl}, G.~I. 1979, in NOAA Solar-Terrestrial Predictions
  Proceedings. Volume 2., ed. R.~F. {Donnelly}, Vol.~2, 258--263

\bibitem[{{Pesnell}(2014)}]{2014SoPh..289.2317P}
{Pesnell}, W.~D. 2014, \solphys, 289, 2317, 2317

\bibitem[{{Roy}(1977)}]{1977SoPh...52...53R}
{Roy}, J.-R. 1977, \solphys, 52, 53, 53

\bibitem[{{Sylwester} {et~al.}(2012){Sylwester}, {Kowalinski}, {Gburek},
  {Siarkowski}, {Kuzin}, {Farnik}, {Reale}, {Phillips}, {Baka{\l}a}, {Gryciuk},
  {Podgorski}, \& {Sylwester}}]{2012ApJ...751..111S}
{Sylwester}, J., {Kowalinski}, M., {Gburek}, S., {et~al.} 2012, \apj, 751, 111,
  111

\bibitem[{{Thompson}(1993)}]{1993SoPh..148..383T}
{Thompson}, R.~J. 1993, \solphys, 148, 383, 383

\bibitem[{{Vaiana} \& {Rosner}(1978)}]{1978ARA&A..16..393V}
{Vaiana}, G.~S., \& {Rosner}, R. 1978, \araa, 16, 393, 393

\bibitem[{{Wolf}(1852)}]{1852MNRAS..13...29W}
{Wolf}, M. 1852, \mnras, 13, 29, 29

\end{thebibliography}

\newpage

\begin{table}[ht]
\caption{Results from Gaussian fits to the 1--8 \AA~X-ray background from GOES using a variety of binning widths.  The solar cycle, peak flux, solar maximum date and the corresponding decimal years since the beginning of the solar cycle (SCY), half-width ($\sigma$) of the solar cycle, date of the end of the cycle, and $\chi^2$ from the model fit are given. Cases where reduced $\chi^2 >> 1$ are indicated as ``oversampled".}\label{table-binning}
\vspace{0.25cm}
\begin{center}
\begin{tabular}{c c c c c l}
\hline\hline
Solar Cycle & $F_0$ & Solar Max & $\sigma$ & End Cycle & $\chi^2$ \\
&	log W m$^{-2}$ & (SCY) & years \\
\hline

1 month	\\
22 &	-5.95 	&  Dec 1990 (4.27)	&6.35	&Apr 1997	&22.2/94 \\
23	&-6.16	&  Sep 2001 (5.38)	&7.69	&Jun 2009	&24.4/118 \\
24	&-6.28	& Jun 2014 (5.53)	&7.75	&Mar 2022	&7.4/35 \\
\hline

2 week			\\
22	&-5.96	& Dec 1990 (4.27)	&6.39	&Apr 1997	&82.7/189 \\
23	&-6.19	& Sep 2001 (5.36)	&8.05	&Oct 2009	&99.2/239 \\
24	&-6.29	& Jan 2014 (5.11)	&6.94	&Dec 2020	&24.0/72 \\
\hline

1 week\\
22	&-5.97	& Dec 12, 1990 (4.28)	&6.40	&May 1997	&326.5/387 \\
23	&-6.21	& Sep 16, 2001 (5.38)	&8.18	&Nov 2009	&386/478 \\
24	&-6.31	& Nov 2013 (5.00)	        &6.76	&Sep 2020	&107.0/153 \\
\hline

0.5 week	\\
22	&-5.98	& Dec 17, 1990 (4.29)	&6.46	&Jun 1997	&1251.2/774 \\
23	&-6.21	& Sep 13, 2001 (5.37)	&8.24	&Dec 2009	&1499.3/956 \\
24	&-6.32	& Oct 2013 (4.82)	&6.25	&Jan 2020	&oversampled \\
\hline

1 day	\\
22	&-5.97	& Dec 17, 1990 (4.29) &	6.46	&Jun 1997	&oversampled \\
23	&-6.21	& Sep 2, 2001 (5.34)	&8.29	&Dec 2009	&oversampled \\
24	&-6.30	& Jun 2013 (4.51)	&5.17	&Aug 2018	&oversampled \\

\hline
\end{tabular}
\end{center}
\end{table}

\begin{table}[ht]
\caption{Results from Gaussian fits to the 1--8 \AA~X-ray background from GOES with the 1-week binning.  The solar cycle, hemisphere-averaged solar maximum flux, hemisphere-averaged solar maximum date and corresponding value in solar cycle years (SCY) or decimal years since the beginning of the solar cycle, and the flux and date of each of the two peaks in each cycle are given.   }\label{table-bestfit}
\vspace{0.25cm}
\begin{center}
\begin{tabular}{c c c c c c c}
\hline\hline
Solar Cycle & $F_0$ & Solar Max  & Peak 1 & Peak 1  & Peak 2 & Peak 2\\
&	log W m$^{-2}$ & (SCY)  &  log W m$^{-2}$ & Date & log W m$^{-2}$ &  Date\\
\hline
22	&-5.97	& Dec 1990 (4.28)	 & -5.59	& Jun 1989	& -5.70	& Apr 1991\\
23	&-6.21	& Sep 2001 (5.38)	& -5.72	& Jul 2000	& -5.87	& Jan 2002\\
24	&-6.31	& Nov 2013 (5.00)	 & -5.82	& Jul 2012	& -5.92	& Feb 2014\\
\hline
\end{tabular}
\end{center}
\end{table}

\begin{table}[ht]
\caption{Details of the monthly sunspot number.  The hemisphere-averaged sunspot (SS) Max values correspond to best-fit median date from a Gaussian fit and the sunspot number at that date.  Peak 1 corresponds to the peak occurring before the fitted median and Peak 2 corresponds to the peak level after the fitted median (SS Max). }\label{table-sunspot}
\vspace{0.25cm}
\begin{center}
\begin{tabular}{c c c c c c c}
\hline\hline
Solar Cycle	&  SS Max & SS Max & Peak 1 & Peak 1 & Peak 2 & Peak 2 \\
&  No. & Date & No. & Date & No. & Date\\
 \hline
22	& 162	 & Jul 1990 & 196 & Jun 1989 & 173 & Jul 1991 \\
23	& 120	 & Mar 2001 & 170 & Jul 2000 & 150 & Sep 2001 \\
24	& 75	 & Nov 2013 & 96 & Nov 2011 & 102 & Feb 2014 \\
\hline
\end{tabular}
\end{center}
\end{table}

\end{document}